\documentstyle[11pt,epsbox,aaspp4]{article}

\slugcomment{YITP-97-56,~KUNS-1480,~astro-ph/9712153}
\lefthead{Sakai, Yokoyama, and Sugiyama}
\righthead{Effect of Void Network on CMB Anisotropy}

\begin{document}

\newcommand{\beq}{\begin{equation}}
\newcommand{\eeq}{\end{equation}}
\newcommand{\bea}{\begin{eqnarray}}
\newcommand{\eea}{\end{eqnarray}}
\newcommand{\etal}{et al.\ }
\newcommand{\gsim}{\gtrsim}
\newcommand{\lsim}{\lesssim}
\newcommand{\bt}{\mbox{\boldmath$\theta$}}
\newcommand{\bdt}{\mbox{\boldmath$\delta\theta$}}
\newcommand{\bda}{\mbox{\boldmath$\delta\alpha$}}
\newcommand{\bk}{\mbox{\boldmath$k$}}
\newcommand{\br}{\mbox{\boldmath$r$}}
\newcommand{\bn}{\mbox{\boldmath$n$}}
\newcommand{\bl}{\mbox{\boldmath$l$}}
\newcommand{\bx}{\mbox{\boldmath$x$}}
\newcommand{\bna}{\mbox{\boldmath$\nabla$}}
\newcommand{\bbv}{\mbox{\boldmath$v$}}

\title{Effect of Void Network on CMB Anisotropy}
\def\thefootnote{\fnsymbol{footnote}}
\affil{
Nobuyuki Sakai$^1$\footnote{E-mail: sakai@yukawa.kyoto-u.ac.jp},
Naoshi Sugiyama$^2$, and Jun'ichi Yokoyama$^{1}$
}
\affil{
$^1$Yukawa Institute for Theoretical Physics, Kyoto University, Kyoto 606-8502, Japan\\
$^2$Department of Physics, Kyoto University, Kyoto 606-8502, Japan\\
}
\begin{abstract}

We study the effect of a void network on the CMB anisotropy in the Einstein-de Sitter background using Thompson \&
Vishniac's model. We consider comprehensively the Sacks-Wolfe effect, 
the Rees-Sciama effect and the gravitational lensing effect.
Our analysis includes the model of primordial voids existing at recombination, which is realized in some
inflationary models associated with a first-order phase
transition. If there exist primordial voids whose comoving radius is larger than
$\sim10h^{-1}$Mpc at recombination, not only the Sachs-Wolfe effect
but also the Rees-Sciama effect is appreciable even for multipoles
$l\lsim1000$ of the anisotropy spectrum.
The gravitational lensing effect, on the other hand, slightly smoothes the
primary anisotropy; quantitatively, our results for the void model are similar to the previous
results for a CDM model. All the effects, together, would give some constraints on the configuration
or origin of voids with high-resolution data of the CMB anisotropy.

\end{abstract}

\keywords{cosmic microwave background, large scale structure}

\section{Introduction}

The anisotropy of the cosmic microwave background (CMB) is an important probe of primordial
fluctuations at recombination, which carries information on the cosmological parameters as well as the
nature of dark matter (see, e.g., \cite{hu97}). CMB photons, however, are also affected gravitationally by
nonlinear structures between recombination and the present epoch. In fact, a network of voids filling the entire
universe has been suggested by redshift surveys such as the CfA2 (\cite{ge89}) and the 
SSRS2 (\cite{da94}). Moreover, using such redshift surveys, El-Ad, Piran \& da Costa (1996,
1997) and El-Ad \& Piran (1997) quantified voids in the galaxy distribution and confirmed the
description of a void-filled universe: they showed that $\sim 50\%$ of the volume is filled with the
voids and that the voids have a diameter of at least $40h^{-1}$Mpc with an average underdensity of
$-0.9$. In this paper we investigate the effect of a void network on the CMB anisotropy in the
Einstein-de Sitter background. 

Rees and Sciama (1968, hereafter RS) showed that an evolving nonlinear structure perturbs the redshift of a
photon passing through it by use of the ``Swiss-cheese" model of overdensity. Later Thompson \& Vishniac
(1987, hereafter TV) estimated this RS effect in a void network model. First, they considered a spherical void in
the Einstein-de Sitter background and derived an analytic expression of the redshift deviation $\delta T/T$
under the thin-shell approximation. Then, using this expression of $\delta T/T$, they calculated the variance of 
$\delta T/T$ for a universe filled with voids. As we will review in \S2, the CMB anisotropy produced by the void
network is of the order $10^{-6}$ if the present diameter of all voids is $60h^{-1}$Mpc and if they have
formed at $z<10$; if the formation time is earlier, the anisotropy becomes larger. These results were supported
by several authors (\cite{sa85}; \cite{mss90}; \cite{ms90}; \cite{me94}; \cite{me96}; \cite{pa92};
\cite{ar93}; \cite{fu96}; \cite{sh96}). Although density perturbations are usually assumed to be
linear at the last-scattering surface (LSS), nonlinear voids can exist there if voids are
originated by primordial bubbles which are nucleated in a phase transition during inflation
(\cite{la91}; \cite{lw91}; \cite{tu92}; \cite{oc94}; \cite{am96}; \cite{bao97}; \cite{bafo97};
\cite{abo98}; \cite{ba98}). Although the hypothesis of primordial voids is quite different from
conventional scenarios, it may explain the present void-network structure and deserves further
consideration. Therefore, more quantitative analysis of the RS effect for that case is important. One of
the purposes of the present analysis is to calculate the power spectrum of the CMB anisotropy by
extending TV's analysis.

The lensing effect of density perturbations, on the other hand, has also been investigated by several authors
(\cite{bl87}; \cite{ka88}; \cite{co89}; \cite{sa89}; \cite{to89}; \cite{tw89};
\cite{fu94}; \cite{li}; \cite{ca}; \cite{se96}; 
\cite{msc97}). It has been concluded, as a whole, that the effect is appreciable on arcminute
angular scales for some models while it is negligible on degree scales. 
In particular, Seljak (1996) solved the shortcomings of the previous studies to
include the nonlinear effects by modeling the power spectrum evolution in the nonlinear regime,
and Mart\'{i}nez-Gonz\'{a}lez, Sanz \& Cay\'{o}n (1997) extended his method to study more
general models; it was shown (\cite{se96}; \cite{msc97}) that for a CDM model the lensing effect
changes the CMB angular power spectrum $C_l$ considerably for
$l\gsim1000$. This power spectrum approach can be applied to
general models as long as the power spectrum of density perturbations is known. Because the power
spectrum of a void-network universe is not obtained, however, we shall estimate the lensing
effect in a different way. That is, we estimate the correction of the
primary anisotropy, using TV's formula of the scattering angle of a photon by a void. An advantage
of our approach is that we make no approximation for nonlinearity nor relativistic effect.

In case nonlinear voids already exist at recombination --- which is the case we are most interested 
in --- we should also include the Sachs-Wolfe (1967, hereafter SW) effect of voids sitting on the
LSS. In fact, it has been investigated by Baccigalupi, Amendola \& Occhionero (1997), Amendola, Baccigalupi \&
Occhionero (1998), and Baccigalupi (1998) to
constrain an inflationary model; it has been shown that the maximum 
radius allowed by the COBE data is $\approx 25h^{-1}$Mpc on the LSS, and that its non-Gaussianity is large enough
to be observable. In their analysis they have ignored the RS term by comparing both terms for a single
void. We agree to their conclusions as a whole, but
it is not clear whether the total contribution is also negligible because the RS effect is generated
not by a single void but by multiple voids between the LSS and us. 
Further, the lensing effect of such voids is also unclear. We 
thus consider the effect of a void network, taking account of the RS 
effect and the lensing effect as well as the SW effect.

The plan of this paper is as follows. In \S2, we briefly review TV's model and results, which we extend in the
following sections. In \S3, we apply the potential approximation to
estimate the SW term. In \S4, we calculate the anisotropy spectra
for the RS effect and for the SW effect of a void network. In \S5, 
we investigate how the primordial fluctuations are modified
by the gravitational lensing effect of a void network.
\S6 is devoted to summary and discussions.

\section{Rees-Sciama Effect --- Thompson \& Vishniac's Model and Results ---}

Because our analysis is based on TV's model of a void-network universe and their analytical results, here we
review them briefly. Consider a single spherical void in the Einstein-de Sitter background:
\beq
ds^2=-dt^2+a^2(t)(dr^2+r^2d\theta^2+r^2\sin^2\theta d\varphi^2) ~~ {\rm with} ~~
a(t)=\Bigl({t\over t_0}\Bigr)^{\frac23},
\eeq
where $t_0$ is the present time.
The void itself is an empty spherical region, and hence a Minkowski spacetime:
\beq
ds^2=-d{t'}^2+d{x'}^2+{x'}^2d\theta^2+{x'}^2\sin^2\theta d\varphi^2,
\eeq
where the prime denotes an internal coordinate. The matter surrounding a void is assumed to form a thin shell.
From momentum conservation and energy conservation, respectively, Maeda \& Sato (1983) and Bertschinger (1985)
showed that the shell radius expands asymptotically as
\beq\label{exlaw}
r_v(t)\propto t^{\beta} ~~ {\rm with} ~~ \beta\approx0.13.
\eeq
Figure 1 shows TV's model of a photon passing through a spherical void. The subscripts 1 and 2 denote
quantities at the time the photon enters the void and at the time it leaves, respectively. We define
$\alpha$ as the angle formed between the direction of observation and the direction of the void's center,
$\delta\alpha$ as the scattering angle of a photon, $d$ as the comoving distance of the void's center, and
$d_{{\rm LSS}}$ as the comoving distance of the LSS.  The angles $\psi_1,~\psi_1',~\psi_2'$ and
$\psi_2$ are defined by reference to Figure \ref{fig1}. 

TV applied double local Lorentz transformations at each void boundary. For example, the relation
between the momentum vector just before entering a void, $\bk_1$, and the one just after passing the
shell, $\bk_1'$, is expressed as
\beq
\bk_1\equiv E_1\left(
\begin{array}{c}
1 \\ \cos\psi_1 \\ \sin\psi_1 \\ 0
\end{array}\right),
\eeq\beq\label{LT}
\bk_1'\equiv E_1'\left(
\begin{array}{c}
1 \\ \cos\psi_1' \\ \sin\psi_1' \\ 0
\end{array}
\right)
=E_1\left(
\begin{array}{c}
\gamma_1\gamma_1'[1+(v_1-v_1')\cos\psi_1-v_1v_1'] \\
\gamma_1\gamma_1'[\cos\psi_1+(v_1-v_1')-v_1v_1'\cos\psi_1] \\
\sin\psi_1 \\ 0
\end{array}\right),
\eeq
where velocities and Lorentz factors are defined as
\beq
v_1\equiv a{dr_v\over dt}\bigg|_{t=t_1},~~~
v_1'\equiv {d(ar_v)\over dt'}\bigg|_{t=t_1},~~~
\gamma_1\equiv{1\over\sqrt{1-v_1^2}},~~~ \gamma_1'\equiv{1\over\sqrt{1-v_1'^2}}.
\eeq
In Appendix A we show the equations of the ``Lorentz transformation" can be derived exactly from
a general coordinate transformation in a curved spacetime. 

After lengthy algebraic calculations, one obtains the redshift deviation and the scattering angle of a photon:
\beq\label{dT}
{\delta T\over T}=(H_2R_2)^3\cos\psi_2\Bigl(3\beta-{2\over3}\cos^2\psi_2\Bigr)
\equiv\Delta_{{\rm RS}},
\eeq
\beq\label{dalpha}
\delta\alpha\equiv-\psi_1+\psi_1'+\psi_2'-\psi_2
=(H_2R_2)^2\sin(2\psi_2),
\eeq
where $H$ is the Hubble parameter, and $R\equiv a r_v$ is the proper length of the shell radius.

Next, TV estimated the CMB anisotropy produced by a void network. Their model consists of 
randomly distributed, equally sized, and non-overlapped voids, which formed at some
time $t_{{\rm f}}$ simultaneously. Divide the universe into shells of the comoving thickness
$2r_v(t_0)$, as depicted in Figure 2. For each shell, the probability of a ray
intersecting a void is given by
\beq
P=\frac32 F_0\left({t\over t_0}\right)^{2\beta},
\eeq
where $F_0$ is the fractional volume of space occupied by voids and normalized at the present.
The variance of $\delta T/T$ for each shell is
\beq\label{sigma1}
\sigma_{{\rm RS,shell}}^2 \equiv
\langle\Delta_{{\rm RS}}^{~2}\rangle_{{\rm shell}}
-\langle\Delta_{{\rm RS}}\rangle_{{\rm shell}}^{~2},
\eeq
where
\beq\label{ave}
\langle\Delta_{{\rm RS}}^{~2}\rangle_{{\rm shell}}=
P \int^{{\pi\over2}}_0\Delta_{{\rm RS}}^{~2} \sin\psi_2 d\psi_2,
\eeq
and $\Delta$ is given by equation (\ref{dT}). Strictly speaking, the integration should be performed with
respect to $\alpha$ instead of $\psi_2$. However, unless a void is very close to an observer,
$\psi_2$ is almost proportional to $\alpha$, and hence the integration (\ref{ave}) is a good
approximation.

The net variance is 
\beq\label{net}
\sigma_{{\rm RS}}^2=\sum\sigma_{{\rm RS,shell}}^2, 
\eeq
where the sum is over all the shells from $t_{{\rm f}}$ to $t_0$. 
We can approximate it with an integral
\beq
\sigma_{{\rm RS}}^2=\int^{t_0}_{t_f}\sigma_{{\rm RS,shell}}^2{dt\over\Delta t}
~~~ {\rm with} ~~~ \Delta t\equiv 2a(t)r_v(t_0).
\eeq
After an algebraic calculation, we obtain
\bea\label{sigma}
\sigma_{{\rm RS}} &=& \frac32(H_0R_0)^{\frac52}\Biggl[{3F_0\over5-24\beta}
\left({2\over81}-{8\over27}\beta+\beta^2\right)
\left\{(1+z_{{\rm f}})^{\frac52-12\beta}-1\right\} \nonumber\\
&&-{4F_0^2\over5(1-6\beta)}\left(\beta-{2\over15}\right)^2
\left\{(1+z_{{\rm f}})^{\frac52-15\beta}-1\right\}\Biggr]^{\frac12}.
\eea
TV mentioned that the second term in equation (\ref{sigma}) goes to
zero if $\beta=2/15$ ($\approx0.133$), or
equivalently, $\langle\Delta_{{\rm RS}}\rangle=0$. Although we take Maeda \& Sato's result $\beta=0.13$
throughout this paper, $\langle\Delta_{{\rm RS}}\rangle\approx0$ is still satisfied.

In the present model there are three parameters: the radius of voids $R_0$, the volume fraction $F_0$, and the
formation time $z_{{\rm f}}$. $z_{{\rm f}}$ should be $1\sim10$ in conventional scenarios of structure formation,
while $z_{{\rm f}}=z_{{\rm LSS}}\approx1000$ in the scenario that voids are originated by primordial
bubbles at the inflationary era. As for $R_0$ and $F_0$, we consider three cases:
\def\theenumi{(\roman{enumi})}
\begin{enumerate}
\item $R_0=20h^{-1}$Mpc and $F_0=60\%$.
\item $R_0=30h^{-1}$Mpc and $F_0=60\%$.
\item $R_0=60h^{-1}$Mpc and $F_0=3\%$.
\end{enumerate}
It must be noted that voids are not at rest in terms of the comoving coordinates but expands
with the power law in equation (\ref{exlaw}). Therefor, for example, a void with $R_0=60h^{-1}$Mpc
at present corresponds to a void with the comoving radius $r_v(t_{{\rm LSS}})\approx15h^{-1}$Mpc on
the LSS.
Models (i) and (ii) are based on the analysis of redshift surveys by El-Ad \etal (1996, 1997) and El-Ad \& and
Piran (1997), which we introduced in \S1. In the real Universe, however, voids should have a smooth distribution
function in size, and a small number of much larger voids may affect the CMB anisotropy. Therefore, we also
consider Model (iii), as an example. Later our result (in Fig.
\ref{fig8}(c))
will show that Model (iii) is compatible with COBE's data. As we
mentioned in \S1, Baccigalupi, Amendola \& Occhionero (1997) also
considered such large voids, and showed that the maximum radius is $\approx25h^{-1}$Mpc on the LSS,
which corresponds to $\approx 100h^{-1}$Mpc at present, from the compatibility between the SW effect
of voids and COBE's data.

To illustrate the typical values of $\sigma_{{\rm RS}}$, we draw a
plot of equation (\ref{sigma}) in Figure \ref{fig3}. We see that the net variance
depends strongly on the formation time of voids. As TV concluded, the RS effect cannot make a
dominant contribution to the CMB anisotropy if nonlinear voids form at $z<10$. On the other hand, 
if voids form before or just after recombination (i.e., $z_{{\rm
f}}\approx1000$), the RS effect may not be negligible.

\section{Sachs-Wolfe Effect --- Estimate with Potential Approximation ---}

As we mentioned in the introduction, for the case where the primordial 
voids exist already at recombination, the SW
effect by those voids should be taken into account. First, we estimate the SW effect of a single
void, using the ``potential approximation" devised by Mart\'{i}nez-Gonz\'{a}lez \etal (1990).

Even if the density profiles are nonlinear, under some condition the metric perturbations in
the Einstein-de Sitter background are characterized by a single potential $\phi(t,\bx)\ll1$:
\beq
ds^2=-(1+2\phi)dt^2+(1-2\phi)a^2(t)(dr^2+r^2d\theta^2+r^2\sin^2\theta d\varphi^2),
\eeq
and the energy-momentum tensor by the matter density $\rho(t,\bx)$ and the velocity field
$\bbv(t,\bx)$. Then one of the Einstein equations reduces to the Poisson equation:
\beq\label{Poi}
{1\over a^2}\bigtriangleup\phi=4\pi G\rho_b{\delta\rho\over\rho},
\eeq
where $\rho_b$ is the background density and $\delta\rho/\rho$ is the density fluctuation
field: $\rho=\rho_b(1+\delta\rho/\rho)$. Mart\'{i}nez-Gonz\'{a}lez \etal (1990) derived the general
expression of redshift fluctuations:
\beq
{\delta T\over T}={1\over3}(\phi_{{\rm LSS}}-\phi_0)-2\int^0_{{\rm LSS}}d\bx\cdot\bna\phi
+\bn\cdot(\bbv_0-\bbv_{{\rm LSS}}).
\eeq
The first, the second, and the third terms are interpreted as the SW effect, the RS effect, and the
Doppler effect, respectively. They also showed that, for an empty void, the second term results in
equation (\ref{dT}), which is TV's result.

Let us calculate the potential inside a void.
For a spherical void with a thin shell, $\rho(t,\bx)$ is explicitly written as
\beq
\rho(t,r)=\rho_b(t)\theta(r-r_v(t))+\rho_{{\rm in}}(t,r)\theta(r-r_v(t))
+\sigma(t)\delta_{{\rm Dirac}}(r-r_v(t)),
\eeq
where $\theta$ is the Heviside function, $\delta_{{\rm Dirac}}$ is Dirac's delta function, 
$\rho_{{\rm in}}$ is the energy density inside the void, and $\sigma$ is the surface energy
density of the shell. If we assume $\rho_{{\rm in}}$ to be homogeneous, 
the Poisson equation (\ref{Poi}) is easily integrated as
\beq\label{SWphi}
\phi={1\over 4}H^2a^2(r^2-r_v^2){\delta\rho\over\rho}, ~~ {\rm for} ~~ 
r<r_v,
\eeq
and $\phi=0$ for $r>r_v$. This result is also obtained in the usual linear perturbation theory. What
we want to emphasize here is, however, that equation (\ref{SWphi}) can
also be applied to nonlinear density profiles such as the present void model. For an empty void
($\delta\rho/\rho=-1$ for $r<r_v$), assuming $\phi_0=0$, we obtain the 
temperature distortion by the SW effect,
\beq\label{dTSW1}
{\delta T\over T}={1\over12}H^2a^2(r^2_v-r^2)|_{{\rm LSS}},
\eeq
which takes a maximal value at $r=0$, corresponding to the 
case where the void's center is just located on the LSS, and vanishes for 
$r>r_v$. In order to take an average over the the location of voids within 
the farthest shell, we define $X$ as the distance between the 
void's center and the LSS (see Fig. 4) and rewrite equation (\ref{dTSW1}) as
\beq\label{dTSW2}
{\delta T\over T}={1\over 12}H^2a^2(r^2_v\cos^2\psi_2^2-X^2) \equiv\Delta_{{\rm SW}}.
\eeq

The variance of $\delta T/T|_{{\rm SW}}$ is calculated as
\beq
\sigma_{{\rm SW}}^{~2} \equiv
\langle\Delta_{{\rm SW}}^{~2}\rangle-\langle\Delta_{{\rm SW}}\rangle^{2},~~
\eeq
where
\beq\label{SWsigma}
\langle\Delta_{{\rm SW}}^{~2}\rangle=
P \int^{{\pi\over2}}_0 \sin\psi_2 d\psi_2 ~
\left[{1\over r_0}\int^{r_v\cos\psi_2}_0 dX ~ \Delta_{{\rm SW}}^{~2}\right].
\eeq
Figure \ref{fig5} shows a plot of equations (\ref{SWsigma}) with a plot of (\ref{sigma}). As long as we look at
the variance of temperature fluctuations, both terms seem to make comparable contributions. The next
task is, of course, to investigate scale-dependent properties of both effects.

\section{Rees-Sciama Effect and Sachs-Wolfe Effect --- Calculation of $C_l$ ---}

For TV's model of a void network, we shall calculate the angular correlation functions of the CMB
anisotropy $C(\theta)$ for the RS term and for the SW term, and their multipole moments
$C_l$, which are defined by
\beq\label{C}
C(\theta)\equiv\left<{\delta T\over T}(\bt_A){\delta T\over T}(\bt_B)\right>_{\theta}
\equiv \sum_l{(2l+1)C_l\over4\pi} P_l(\cos\theta),
\eeq
where $\bt_A$ and $\bt_B$ are angular positions with the separation angle $\theta$. 


First, let us consider the RS effect. In general, in order to calculate the correlation function with
density configuration in a real space, simulation-like computation is needed. Once we evaluate $C_{{\rm
RS}}(\theta)$ for each shell, however, we can sum up the contributions from all shells by using the same
relation as equation (\ref{net}). Furthermore, for each shell, by the assumption of random distribution
of voids and by the relation $\langle\Delta_{{\rm RS}}\rangle\approx0$, the correlation function
$C_{{\rm RS}}(\theta)$ for the case where two light rays $A$ and $B$ pass different voids
becomes zero; this allows us to consider only one void in our calculation. Let us imagine a void
projected on the celestial sphere, as depicted in Figure \ref{fig6}. The angles
$\alpha_A,~\alpha_B,~\alpha_c,~\alpha_s$, and $\xi$ are defined by reference 
to Figure \ref{fig6}. We
calculate $C_{{\rm RS}}(\theta)$ for each shell by neglecting the curvature of the sphere in 
a local region around a void. Defining the midpoint of $\bt_A$ and $\bt_B$ as 
\beq
\bt_c\equiv{\bt_A+\bt_B\over2}=(\alpha_c,~0),
\eeq
the positions of $\bt_A$ and $\bt_B$ are written as 
\beq\label{btab}
\bt_A=\left(\alpha_c+{\theta\over2}\cos\xi,~{\theta\over2}\sin\xi\right), ~~~
\bt_B=\left(\alpha_c-{\theta\over2}\cos\xi,~-{\theta\over2}\sin\xi\right).
\eeq
Then we can calculate $C_{{\rm RS}}(\theta)$ for each shell with the expression 
\beq\label{Crs}
C_{{\rm RS,shell}}(\theta)
= {\alpha_{{\rm m}}^2\over\alpha_{{\rm s}}^2}P ~
{2\over\alpha_{{\rm m}}^2}\int^{\alpha_{{\rm m}}}_0 \alpha_c d\alpha_c
\left[{1\over\pi}\int^{\pi}_0 d\xi
\Delta_{{\rm RS}}(\alpha_A)\Delta_{{\rm RS}}(\alpha_B)\right],
\eeq
where the relation between $\psi_2$ and $\alpha ~(\alpha_A$ or $\alpha_B$) is
\beq\label{geo1}
\sin\psi_2={d\over r_2}\sin\alpha ~~~ {\rm with} ~~~
d=3t_0\left[1-\left({t\over t_0}\right)^{\frac13}\right], 
\eeq
and the angular size $\alpha_{{\rm s}}$ which corresponds to the void radius and the integral boundary
$\alpha_{{\rm m}}$ are defined as
\beq
\alpha_{{\rm s}}\equiv\alpha|_{\psi_2={\pi\over2}}=\arcsin\left({r_2\over d}\right), ~~~
\alpha_{{\rm m}}\equiv\alpha_{{\rm s}}+{\theta\over2}.
\eeq

The correlation function for the SW effect is calculated similarly: $C_{{\rm SW}}(\theta)$ is given by
\beq\label{Csw}
C_{{\rm SW}}(\theta)
={\alpha_{{\rm m}}^2\over\alpha_{{\rm s}}^2}P ~
{2\over\alpha_{{\rm m}}^2}\int^{\alpha_{{\rm m}}}_0 \alpha_c d\alpha_c
\left\{{1\over\pi}\int^{\pi}_0 d\xi ~
\left[{1\over r_0}\int^{X(\Delta_{{\rm SW}}=0)}_0 dX
\Delta_{{\rm SW}}(\alpha_A)\Delta_{{\rm SW}}(\alpha_B)\right]\right\},
\eeq
where $\Delta(\alpha)$ has been redefined as $\Delta(\alpha)-\langle\Delta(\alpha)\rangle\longrightarrow
\Delta(\alpha)$ so that $\langle\Delta(\alpha)\rangle=0$.

Numerical integration of equations (\ref{Crs}) and (\ref{Csw}) gives us $C_{{\rm RS}}(\theta)$ and
$C_{{\rm SW}}(\theta)$, and their multipole moments are obtained from
\beq
C_l=2\pi\int^1_{-1}C(\theta)P_l(\cos\theta)d(\cos\theta),
\eeq
which are equivalent to equation (\ref{C}).

Figure \ref{fig7} reports how the anisotropy spectrum of the RS effect depends on the formation time $z_{{\rm
f}}$. Here we plot $\sqrt{l(l+1)C_l}$ versus $\log l$ in accordance with Amendola, Baccigalupi \&
Occhionero (1998). As we expected, the RS effect for $z_{{\rm f}}=10$ is negligibly small while it is
not for $z_{{\rm f}}=1000$. 

In Figure \ref{fig8} we compare the RS effect and the SW effect for our three models.
In Model (i) the RS effect is negligibly small; in Model (ii) it is still smaller
than the SW effect for $l<1000$, but it is more than $10\%$ of the SW term 
for $l\lsim 1000$ and not negligible; in Model (iii) both terms are
comparable. The dependence of $C_l$ on $F_0$ is easily understood: as equation
(\ref{sigma}) indicates, $\delta T/T$ or $C_l$ is proportional to $\sqrt{F_0}$.

\section{Gravitational Lensing Effect}

We now turn to the lensing effect. First, we estimate the characteristic
angular scale below which the lensing effect is appreciable, by calculating
the angular excursion of a photon on the LSS, $\bdt$. The lens equation for a
single void in the thin lens approximation is
\beq\label{lenseq}
\bdt_{{\rm one~void}}=W\delta\alpha\bl ~~~ {\rm with} ~~~
W\equiv{d_{{\rm LSS}}-d\cos\alpha\over d_{{\rm LSS}}},
\eeq
where the vector $\bl\equiv\bdt/|\bdt|$ has been introduced in
Figure 1. In reality this thin lens equation can be derived exactly from null geodesic  equations, as
shown in Appendix B.

Replacing $\delta T/T$ with $W\delta\alpha$ in TV's calculation of $\sigma_{{\rm RS}}$ in \S2, we find
$$
\sqrt{\langle|\bdt|^2\rangle} = (H_0R_0)^{\frac32}{\sqrt{F_0}\over\sqrt{z_{{\rm LSS}}+1}-1}
\Biggl[ {1\over3(6\beta-1)}-{\sqrt{z_{{\rm LSS}}+1}\over9\beta-1}+{z_{{\rm LSS}}+1\over18\beta-1}
$$
\beq\label{var}
-(z_{{\rm f}}+1)^{\frac32-9\beta}
\left\{{1\over3(6\beta-1)}-\sqrt{{z_{{\rm LSS}}+1\over z_{{\rm f}}+1}}{1\over9\beta-1}
+{z_{{\rm LSS}}+1\over z_{{\rm f}}+1}{1\over18\beta-1}\right\}
\Biggr]^{\frac12}.
\eeq
We draw a plot of equation (\ref{var}) in Figure \ref{fig9}, assuming $z_{{\rm LSS}}=1000$. The lensing
effect also depends on the formation time of voids, though the dependence is not so strong as in the
case of the RS effect. The typical angular scale of lensing is several arcminutes.

These angles, however, are not directly observable; we would rather calculate the
dispersion of the relative angular separation $\theta$, which is defined as
\beq
\sigma_{{\rm gl}}(\theta)\equiv
{1\over\sqrt2}{\langle|\bdt_A-\bdt_B|^2\rangle_{\theta}}^{\frac12}.
\eeq
$\sigma_{{\rm gl}}$ is, like the case of the RS effect, proportional to $\sqrt{F_0}$. The values of
$\sigma_{{\rm gl}}(\theta)$ for each shell and their sum are computed just like the computation of
$C_{{\rm RS}}(\theta)$ in \S3. Figure \ref{fig10} shows a plot of $[\sigma_{{\rm gl}}(\theta)/\theta]_{\theta=0.01
{\rm arcmin}}$ for each shell. Voids at $z\approx50$ make the largest contribution to the CMB anisotropy:
due to the factor $W$ in equation (\ref{lenseq}), no effect arises from voids on the LSS, contrary to the
case of the RS effect.

The values of $\sigma_{{\rm gl}}(\theta)/\theta$ for the three models are 
presented in Figure \ref{fig11}. The results for
Models (i) and (ii) indicate that the dependence on the formation time $z_{{\rm f}}$ is not so strong,
as above. We also note that the results are similar to the previous results for a CDM model
(\cite{se96}; \cite{msc97}), though models and methods are quite different. The lensing effect for
Model (iii) is not so large as that in Model (i) or (ii), though the RS effect is
maximal in Model (iii) as shown in Figure. \ref{fig8}(c). This difference stems
from the dependence of each effect on the void scale: as equations
(\ref{sigma}) and (\ref{var}) shows, $\delta T/T|_{{\rm RS}}$ is proportional to
$(H_0R_0)^{5/2}$, while $\sqrt{\langle|\bdt|^2\rangle}$ is proportional to $(H_0R_0)^{3/2}$.

Once $\sigma_{{\rm gl}}(\theta)$ is obtained, we can calculate the lensing effect on the CMB
fluctuations. Here we adopt the approximate formula of Seljak (1996) for the lensed correlation function:
\beq
\tilde C(\theta)={1\over2\pi}\int^{\infty}_0ldle^{-\sigma_{{\rm gl}}^{~2}(\theta)l^2/2}C_{l}J_0(l\theta).
\eeq
An example of the CMB anisotropy spectrum including lensing is presented 
in Figure \ref{fig12}(a). Here we plot $l(l+1)C_l$ versus $l$ in 
linear scales in accordance with Seljak (1996). As Seljak (1996) and 
Mart\'{i}nez-Gonz\'{a}lez \etal (1997) showed for a CDM model, the lensing 
effect is to slightly smooth the main features appearing in the spectrum.
To see the dependence on $R_0$ and $z_{{\rm f}}$, we show the relative 
changes of the spectrum due to lensing for several cases in Figure \ref{fig12}(b), (c).
We find that the lensing effect has a weak dependence on the formation time. That is, even if nonlinear voids exist
from the recombination, they do not change the CMB anisotropy significantly. This feature is in contrast with that
of the RS effect.

\section{Summary and Discussions}

We have studied the effect of a void network on the CMB anisotropy for TV's model, where many
voids are distributed in the Einstein-de Sitter background. In particular, we have examined how the CMB anisotropy
spectrum affected by the RS effect, the SW effect, and the gravitational lensing effect.

Although the RS effect is negligible in conventional scenarios of structure formation like a CDM model, it can be
appreciable if primordial voids exist already at recombination. In such a case, the SW effect of voids lying on the LSS is
also important, and hence we have compared the two effects for several models. In most cases the SW 
term is larger than the RS term for $l\lsim 1000$; however, if there are voids with the present
radius $R_0\gsim 60h^{-1}$Mpc, or equivalently, $r_v(t_{{\rm LSS}})\gsim 
15h^{-1}$Mpc on the LSS, both effects are comparable. Moreover, RS is 
the dominant effect on small scales.

For the SW effect of voids, Amendola, Baccigalupi \& Occhionero (1998) 
and Baccigalupi (1998) argued that non-Gaussianity is large and it may give
rise to ordered patterns in the CMB anisotropy field. For the RS effect, on the other hand, the
deviation from Gaussianity depends on the number density of voids, or the volume fraction $F_0$. If
$F_0$ is so small as in Model (iii), the main contribution to the RS effect is made by a few voids
near the LSS; then similar characteristic patterns may appear on the CMB map. On the other hand,
if $F_0$ is of order unity as in Model (i) or (ii), the RS effect is generated by many (typically,
more than ten) voids. In this case the central limit theorem implies that 
Gaussian statistics is a good approximation for the RS term itself; however, non-Gaussianity
of the SW term exists and is dominant. It needs further investigation 
to clarify Gaussian/non-Gaussian feature of the RS effect more quantitatively.

As for gravitational lensing, the effect of nonlinear voids has not been investigated so far. Our
present work is the first trial of that investigation without any approximation for nonlinearity nor
relativistic effect. We have shown how the primary anisotropy is smoothed out for some values of the
void radius $R_0$ and of the formation time $z_{{\rm f}}$. We have found that our results are
similar to those for a CDM model (\cite{se96}; \cite{msc97}). 

Our results as a whole suggest that, if the real universe is filled with nonlinear voids, they make some
appreciable effects on the CMB anisotropy. Those effects are
expected to give some constraints on the configuration or origin of voids, particularly on the inflationary
models in which primordial bubbles are nucleated, with the next generation of CMB satellites.

\section*{Acknowledgments}

We thank Paul Haines for reading the manuscript. 
Numerical Computation of this work was carried out at the Yukawa Institute Computer Facility. N. Sakai was
supported by JSPS Research Fellowships for Young Scientists. This work was supported partially by the
Grant-in-Aid for Scientific Research Fund of the Ministry of Education, Science and Culture (No.9702603,
No.09740334, and No.09440106).

\appendix
\section{On Double Lorentz Transformation}

In this Appendix we show the equations of the double ``Lorentz transformation" (\ref{LT}) can be derived
exactly from a general coordinate transformation in a curved spacetime. Here we demonstrate the
transformation at $t=t_1$ and omit the subscript 1.

First, we have to introduce another coordinate system which overlaps both the Einstein-de Sitter
background and the Minkowski region. We adopt a Gaussian normal coordinate system
$(\tau,n,\theta,\varphi)$ in which $n=0$ represents the world-hypersurface of the shell.
$\tau$ is chosen to be the proper time of the shell. i.e., the 3-metric at $n=0$ is give by
\beq
ds^{2}=-d\tau^{2}+R^{2}(\tau)(d\theta^2+\sin^2\theta d\varphi^2). \label{ds3}
\eeq
The coordinate transformation of the 4-momentum $k^{\mu}$ at $n=0$ from the Einstein-de Sitter
frame $\{x^{~~\mu}_{{\rm EdS}}\}$ to the Gaussian normal frame $\{y^{~~\nu}_{{\rm GN}}\}$ is given by
\beq\label{kmu}
k_{~~\mu}^{{\rm GN}}={\partial x^{~~\nu}_{{\rm EdS}}\over \partial y^{~~\mu}_{{\rm GN}}}
k_{~~\nu}^{{\rm EdS}}.
\eeq
It is easy to find
\beq
{\partial t\over\partial\tau}=\gamma,~~~ {\partial r\over\partial\tau}={\gamma v\over a},
\eeq
and Sakai \& Maeda (1993) obtained
\beq
{\partial t\over\partial n}=\gamma v,~~~ {\partial r\over\partial n}={\gamma\over a}.
\eeq
If we write the 4-momentum in each frame as
\beq
k^{~~\mu}_{{\rm EdS}}=E\Bigl(1,~{\cos\psi\over a},~{\sin\psi\over ar},~0\Bigr),~~~
k^{~~\mu}_{{\rm GN}}=E''\Bigl(1,~\cos\psi'',~{\sin\psi''\over R},~0\Bigr),
\eeq
equation (\ref{kmu}) reduces
\beq
E''=E\gamma(1+v\cos\psi),~~~ E''\cos\psi''=E\gamma(v+\cos\psi).
\eeq
This is nothing but the expression of a usual Lorentz transformation. If we also carry out the coordinate
transformation of the 4-momentum from the Gaussian normal frame to the Minkowski frame in the similar way,
the expression of the double Lorentz transformation (\ref{LT}) is obtained.

\section{Derivation of Thin Lens Equation}

Here we derive the lens equation (\ref{lenseq}) from null geodesic equations. As
depicted in Figure 1, we introduce a vector basis, $\{\bn(\alpha),\bl (\alpha)\}$, in the
two-dimensional comoving space, and denote each position by a vector symbol $\br$. The trajectory of a
photon in a homogeneous region is 
\bea\label{r2-r0}
\br_2-\br_0&=&\int^{t_0}_{t_2}{dt\over a}\bn=3t_0(1-\sqrt{a_2})\bn,\\
\label{re-r1}
\br_{{\rm LSS}}-\br_1&=&\int^{t_{{\rm LSS}}}_{t_1}{dt\over a}(\cos\delta\alpha\bn+\sin\delta\alpha\bl)
=3t_0\left[(\sqrt{a_1}-\sqrt{a_{{\rm LSS}}})(\bn+\delta\alpha\bl)+O(\zeta^3)\right],
\eea
where we have expanded the equations with a dimensionless parameter,
\beq
\zeta\equiv {r_2\over3t_0}. 
\eeq
To get the last equality of (\ref{re-r1}), we have used the facts that $\delta\alpha$ is of
order $(H_2R_2)^2$ and that $H_2R_2$ is of order $\zeta$.
Explicitly, the geodesic between $\br_0$ and $\br_2$ says equation (\ref{geo1}) and 
\beq\label{eta2}
H_2R_2={2\zeta\over1-d\cos\alpha/3t_0+\zeta\cos\psi_2},
\eeq
and hence equation (\ref{dalpha}) is rewritten as
\beq\label{dalpha2}
\delta\alpha={8q\sqrt{1-q^2}\over(1-d\cos\alpha/3t_0)^2}\zeta^2+O(\zeta^3) ~~ {\rm with} ~~
q\equiv{d\over r_2}\sin\alpha.
\eeq
\noindent
As for $\br_1-\br_2$, we obtain
\beq\label{r1-r2-1}
\br_1-\br_2=\left[r_1\cos(\delta+\psi_1)+r_2\cos\psi_2\right]\bn
+\left[r_1\sin(\delta+\psi_1)-r_2\sin\psi_2\right]\bl.
\eeq
In order to expand $\delta+\psi_1,~r_1$ and $\sqrt{a_1}$ with $\zeta$ (or $H_2R_2$), we use the
relations
\beq\label{eta1}
H_1R_1=H_2R_2-(3\beta-1)\cos\psi_2(H_2R_2)^2+\frac32(3\beta-1)
\biggl[\Bigl(\beta-\frac23\Bigr)\cos^2\psi_2+\beta\biggr](H_2R_2)^3
+O\left((H_2R_2)^4\right),
\eeq\beq\label{psi1}
\psi_1=\psi_2+3\beta H_2R_2\sin\psi_2+O\left((H_2R_2)^2\right),
\eeq
which are found in the derivation of TV's formula (\ref{dT}) and (\ref{dalpha}). Then we obtain
\beq\label{r1-r2}
\br_1-\br_2=r_2(2\cos\psi_2-3\beta H_2R_2)\bn+3t_0 O(\zeta^3)
=3t_0[(\sqrt{a_2}-\sqrt{a_1})\bn+O(\zeta^3)].
\eeq
Equation (\ref{r1-r2}) tells us that, although a photon is bent inside the void, the
deviation is of order $\zeta^3$. From equations (\ref{r2-r0})-(\ref{eta2}), (\ref{r1-r2}), and the
relation,
\beq
\sqrt{a_1}=\sqrt{a_2}+O(\zeta)=1-{d\cos\alpha\over3t_0}+O(\zeta),
\eeq
we find
\bea
\br_{{\rm LSS}}-\br_0&=&
3t_0\left[(1-\sqrt{a_{{\rm LSS}}})\bn +\left(1-{d\cos\alpha\over3t_0}-\sqrt{a_{{\rm LSS}}}\right)\delta\alpha\bl
+O(\zeta^3)\right]\nonumber\\
&=&(\br_{{\rm LSS}}-\br_0)_b+(d_{{\rm LSS}}-d\cos\alpha)\delta\alpha\bl+3t_0O(\zeta^3),
\label{re-r0}\eea
where the subscript $b$ denotes a background unperturbed quantity. We thus 
reach equation (\ref{lenseq}), i.e., 
\beq
\bdt\equiv{\br_{{\rm LSS}}-\br_{{\rm LSS},b}\over d_{{\rm LSS}}}
={d_{{\rm LSS}}-d\cos\alpha\over d_{{\rm LSS}}}\delta\alpha\bl+O(\zeta^3).
\eeq
This result provides us a simple description: as long as we look at the leading terms of $\zeta$, a
void can be regarded to be a ``thin lens" with a scattering angle $\delta\alpha$.


\clearpage

\epsfile{file=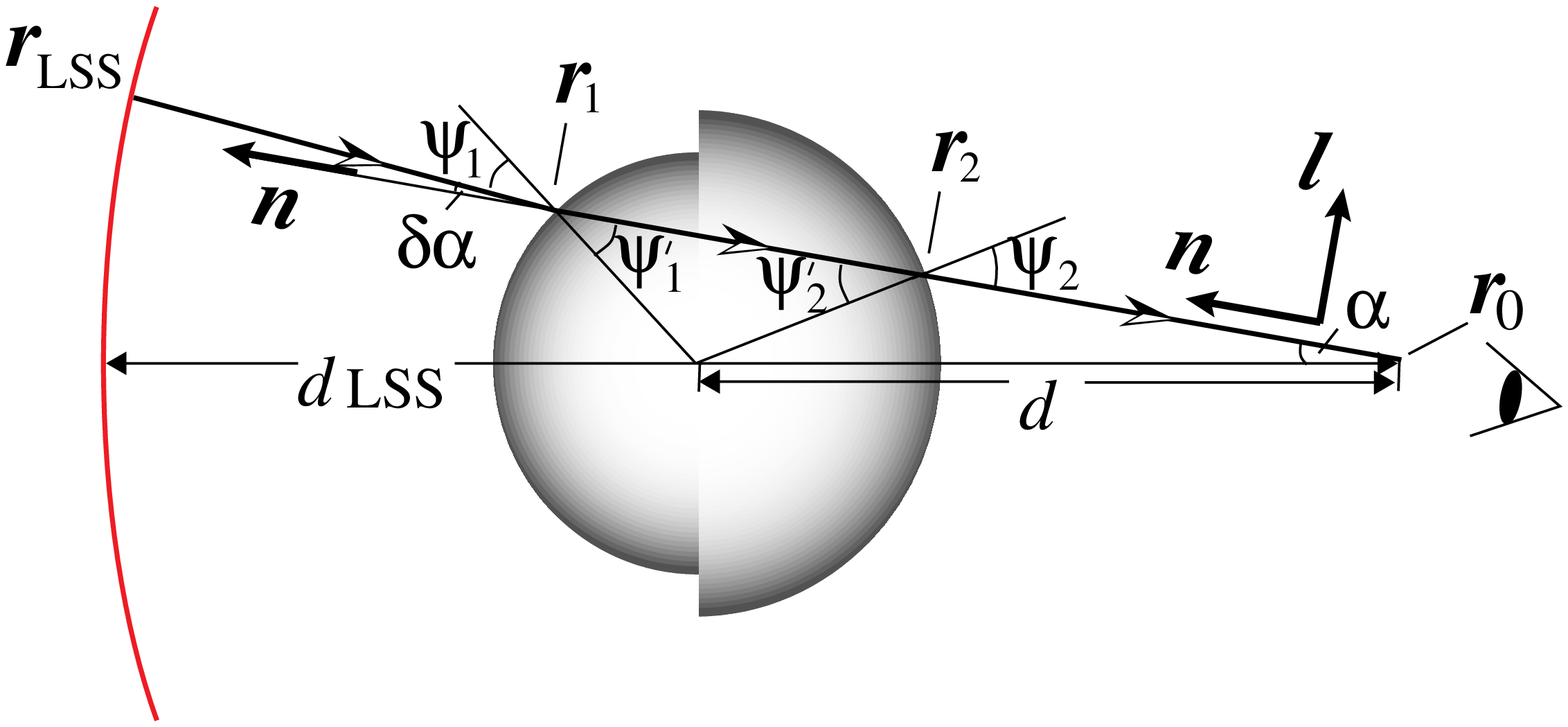,width=12cm}
\figcaption{Cross section of a void on the $\varphi=\pi/2$ plane (TV's model). We depict the trajectory
of a photon from the LSS to an observer. The subscripts 1 and 2 denote quantities at the time the
photon enters the void and at the time it leaves, respectively. The subscripts LSS and 0 denote
quantities at the LSS and at present, respectively. Define $\alpha$ as the angle formed between the
direction of observation and the direction of the void's center, $\delta\alpha$ as the scattering angle of a
photon, $d$ as the comoving distance of the void's center, and $d_{{\rm LSS}}$ as the comoving distance of
the LSS. For Appendix B we denote each position by a vector symbol $\br$, and introduce a vector basis,
$\{\bn(\alpha),\bl (\alpha)\}$, in the two-dimensional comoving space. \label{fig1}}

\vskip 5mm
\epsfile{file=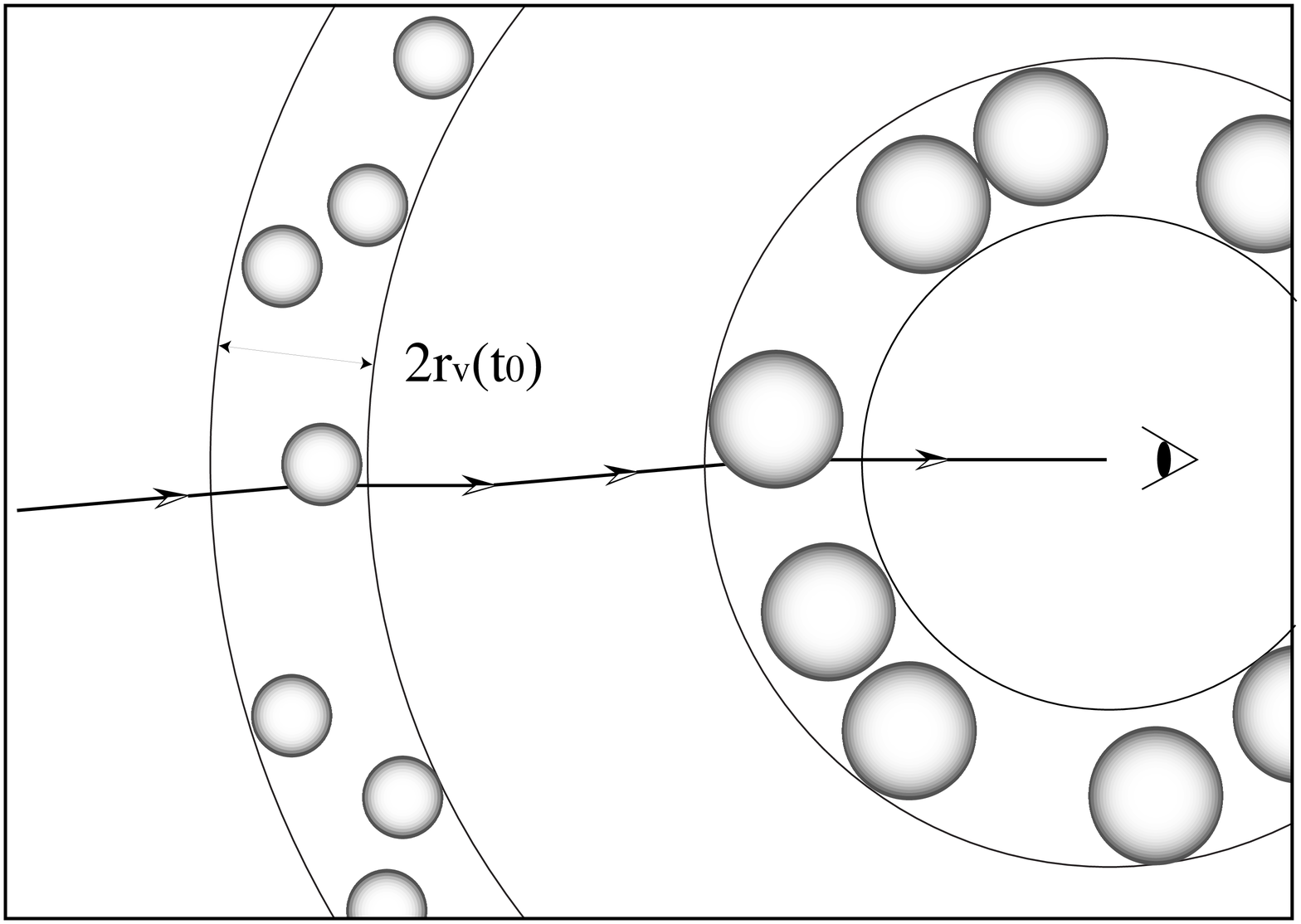,width=11cm}
\figcaption{Schematic sketch of a void-network universe (TV's model). Divide the universe into shells of
comoving thickness $2r_v(t_0)$.
\label{fig2}}

\epsfile{file=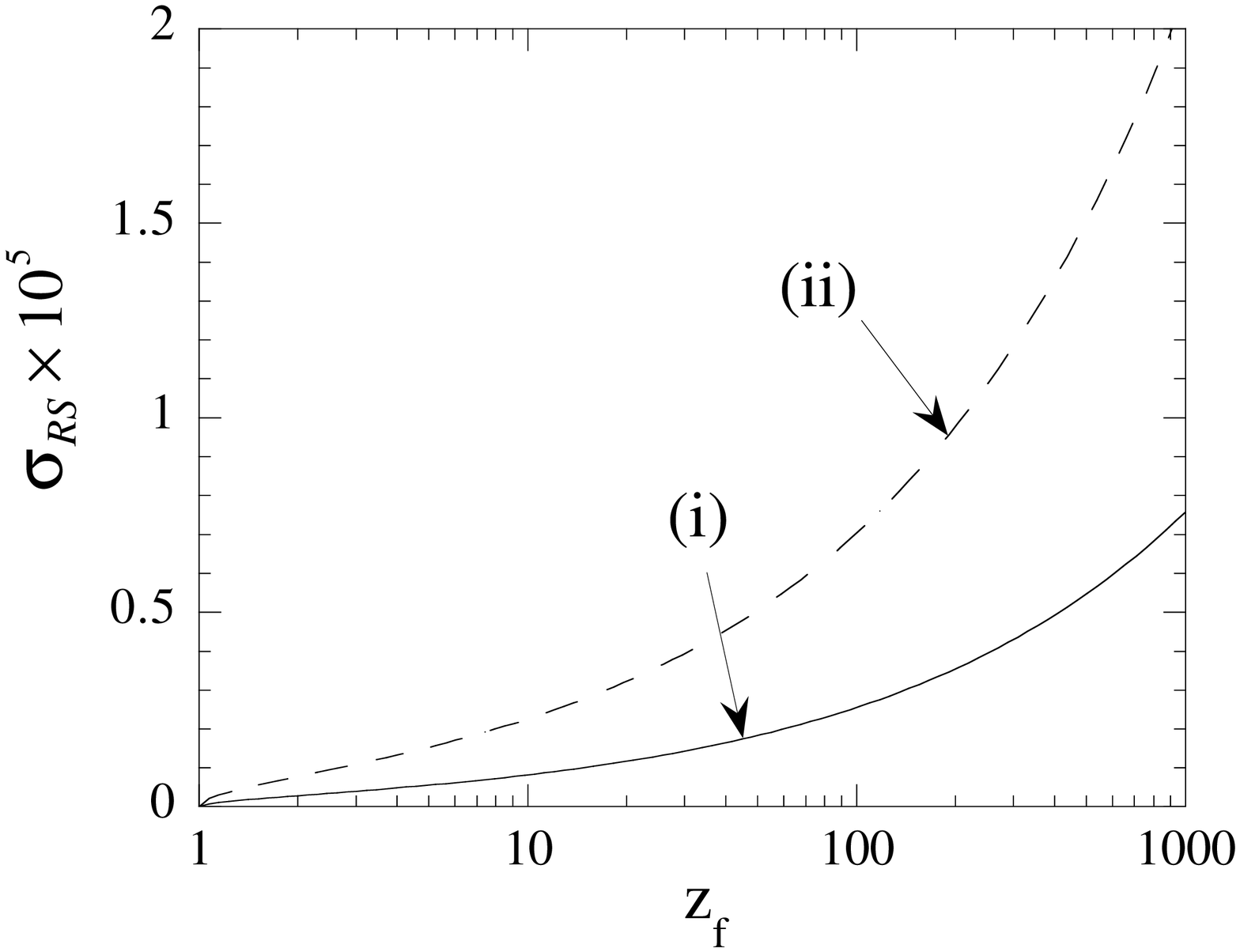,width=12cm}
\figcaption{RS effect: plot of eq. (\ref{sigma}) (TV's results). 
We set $\beta=0.13$ throughout this paper. \label{fig3}}

\vskip 5mm
\epsfile{file=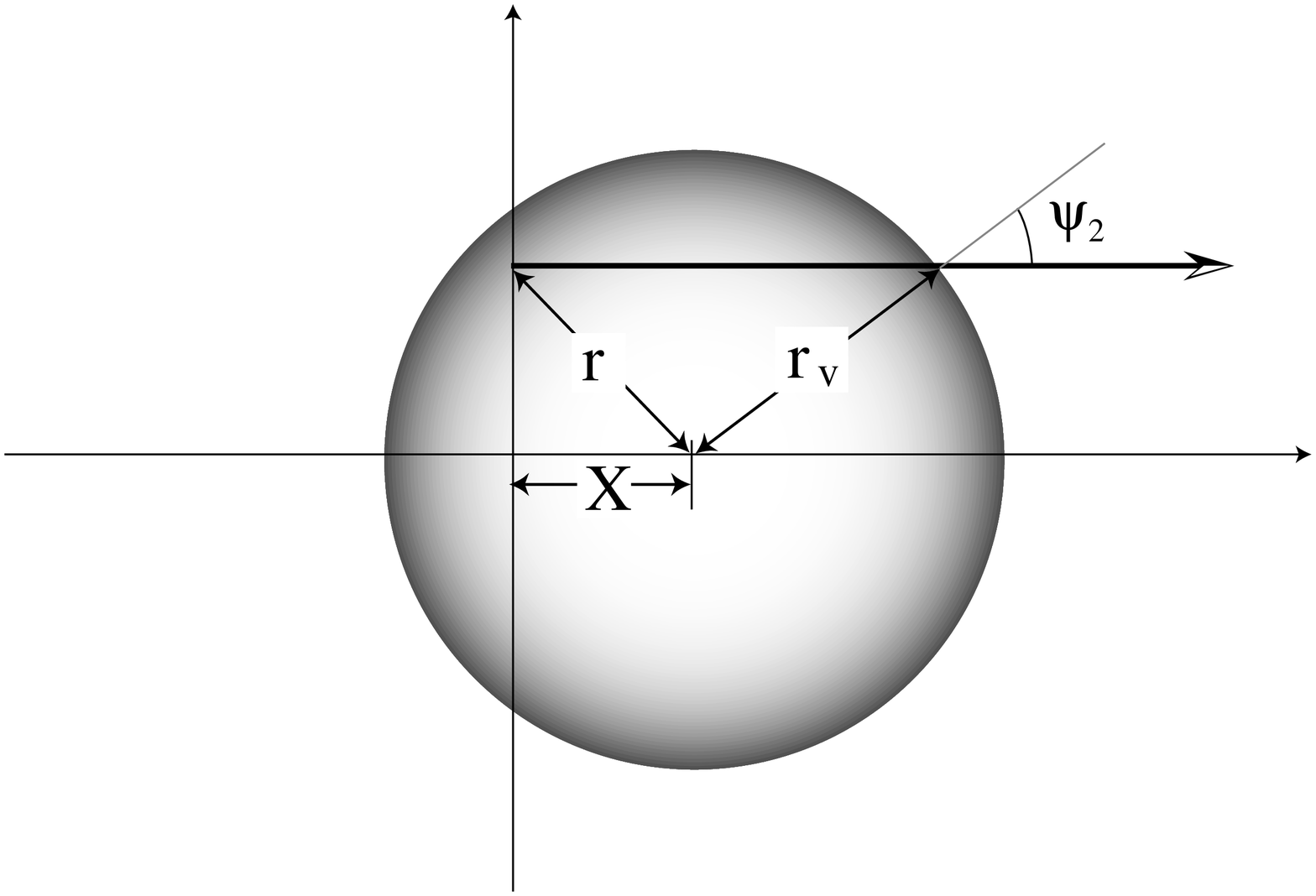,width=11cm}
\figcaption{Void on the LSS. The vertical axis corresponds to the LSS, 
and X is defined as the distance between the void's center and the 
LSS. \label{fig4}}

\epsfile{file=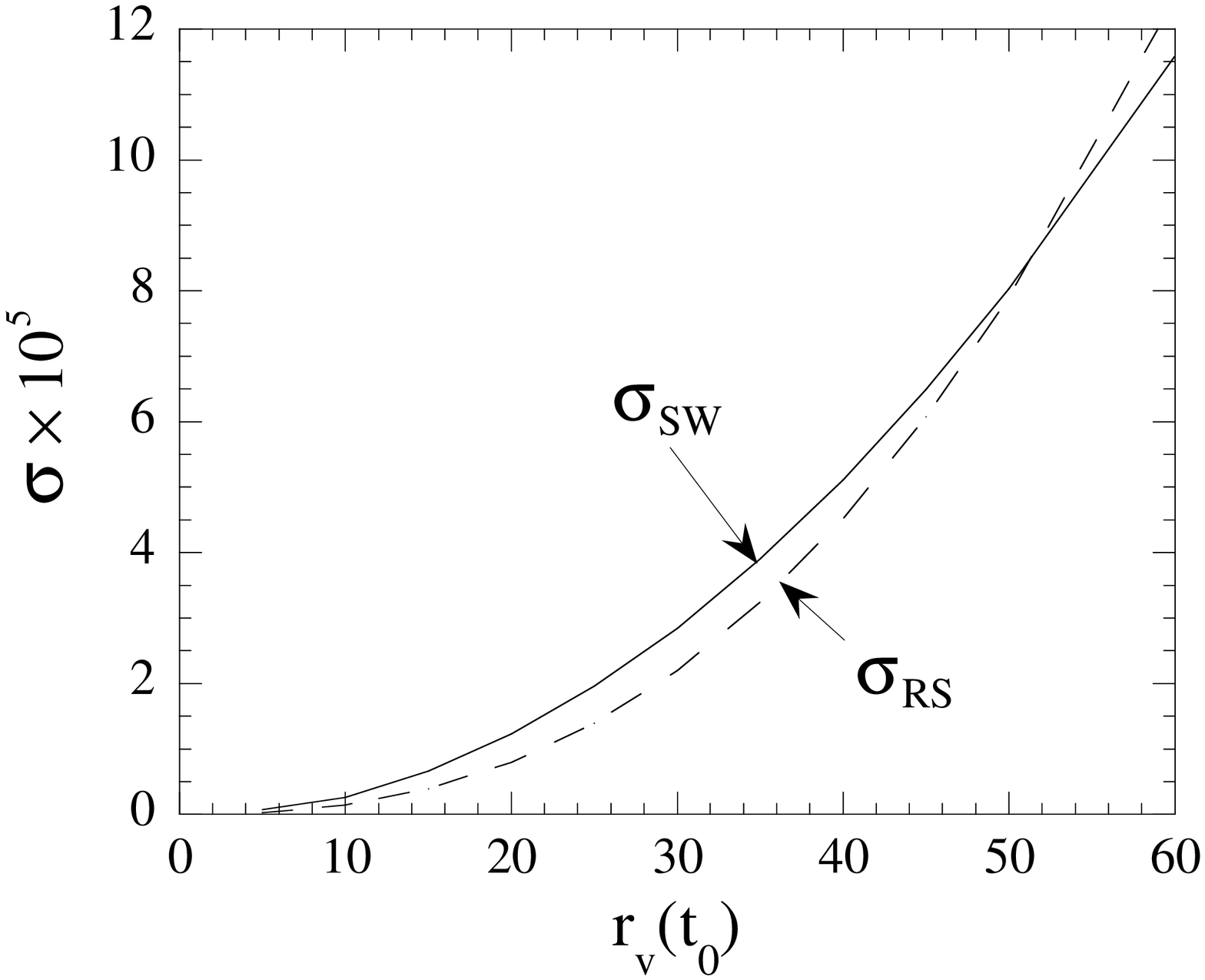,width=12cm}
\figcaption{RS effect and SW effect: plot of $\sigma_{{\rm RS}}$ and $\sigma_{{\rm SW}}$. v.s. $r_v(t_0)$. 
We set $F_0=50\%$. \label{fig5}}

\vskip 5mm
\epsfile{file=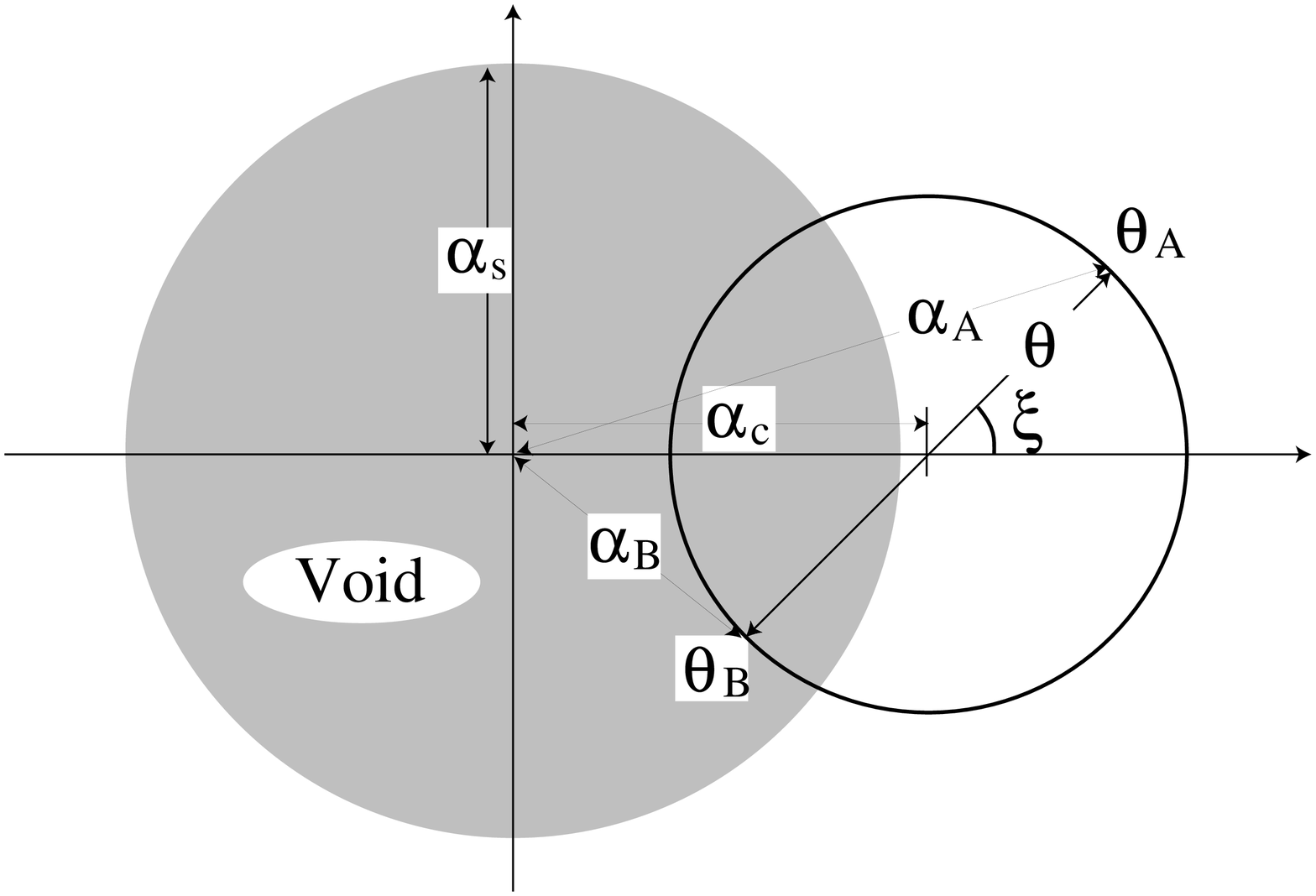,width=11cm}
\figcaption{Projection of a void on the celestial sphere. Neglecting the curvature of the sphere in a local region
around a void, we consider two light rays, which are labeled as $\bt_A$ 
and $\bt_B$. \label{fig6}}

\epsfile{file=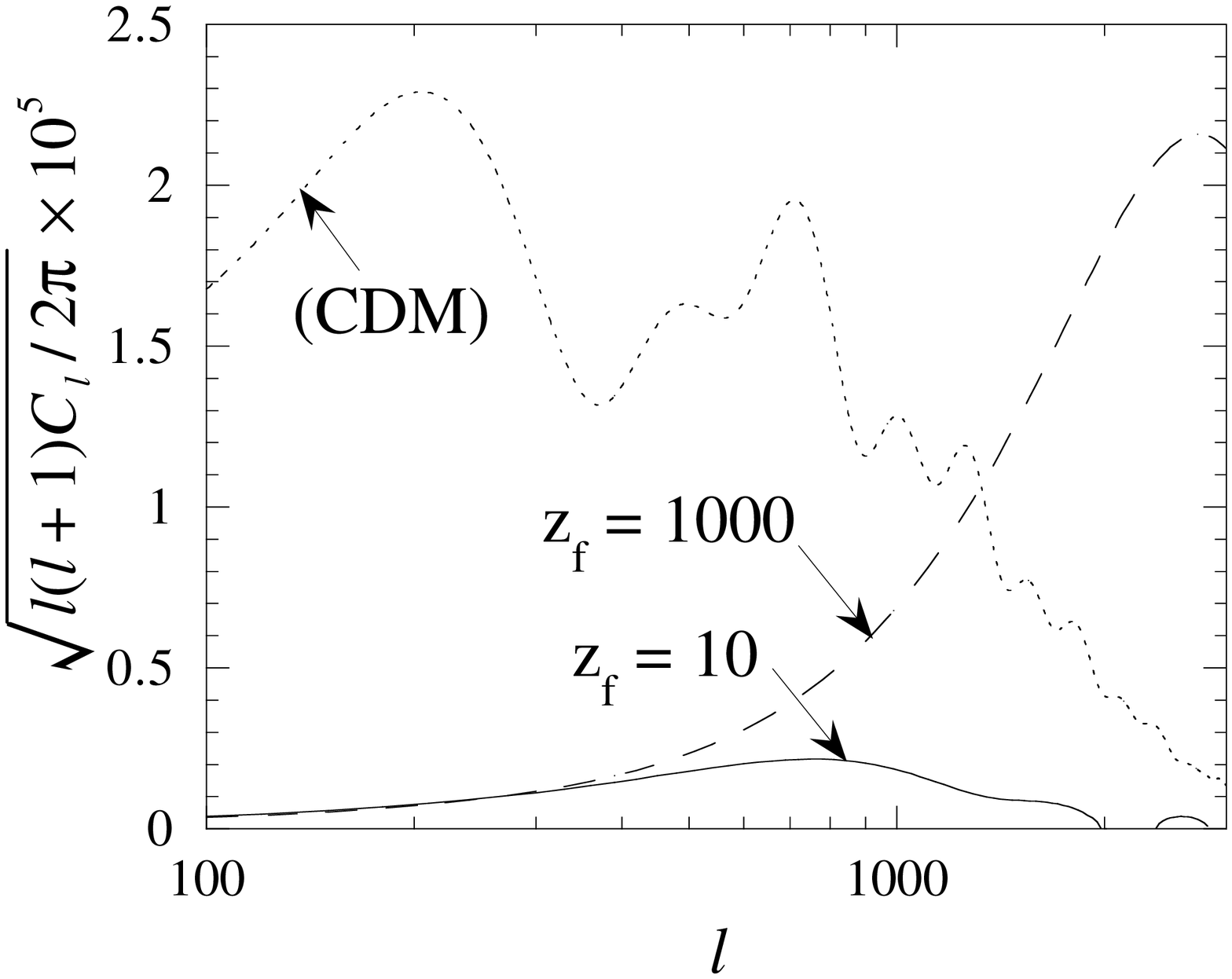,width=12cm}
\figcaption{RS effect: plot of the CMB angular power spectrum $\sqrt{l(l+1)C_l}$
v.s. $l$ for Model (ii). For reference, we also plot the values of $C_l$ estimated for a CDM model.
\label{fig7}}

\vskip 5mm
(a) \epsfile{file=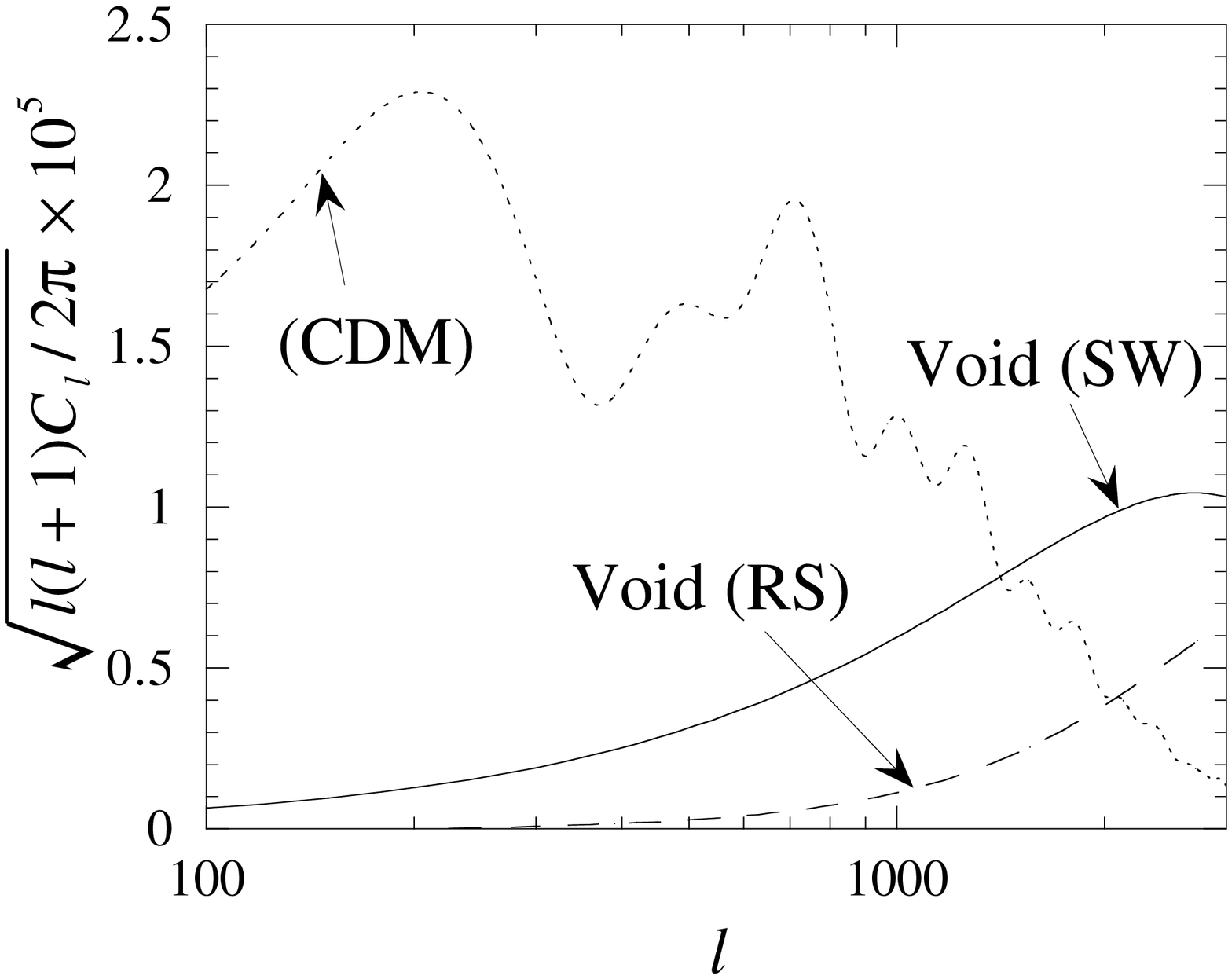,width=12cm}

(b) \epsfile{file=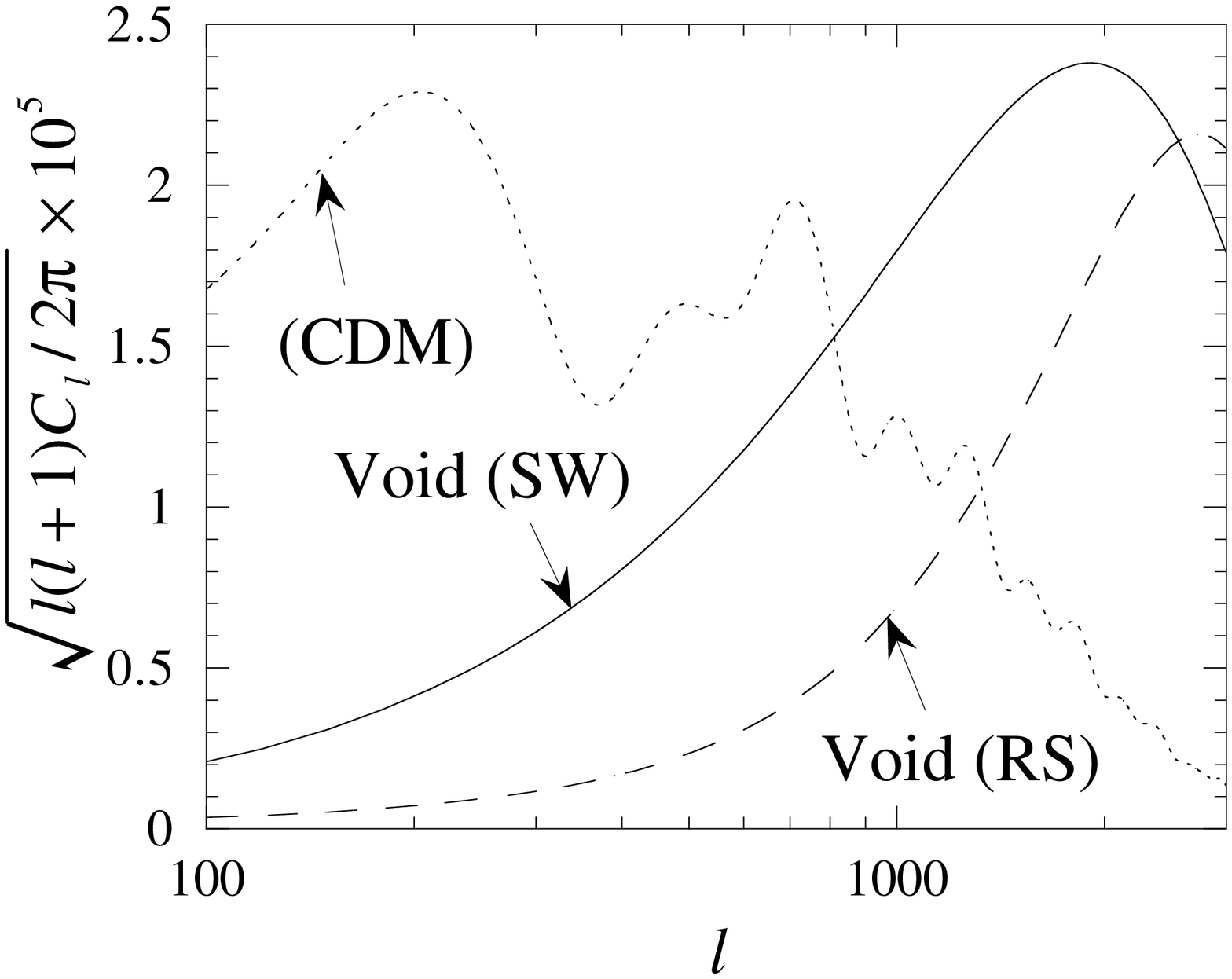,width=12cm}

\vskip 5mm
(c) \epsfile{file=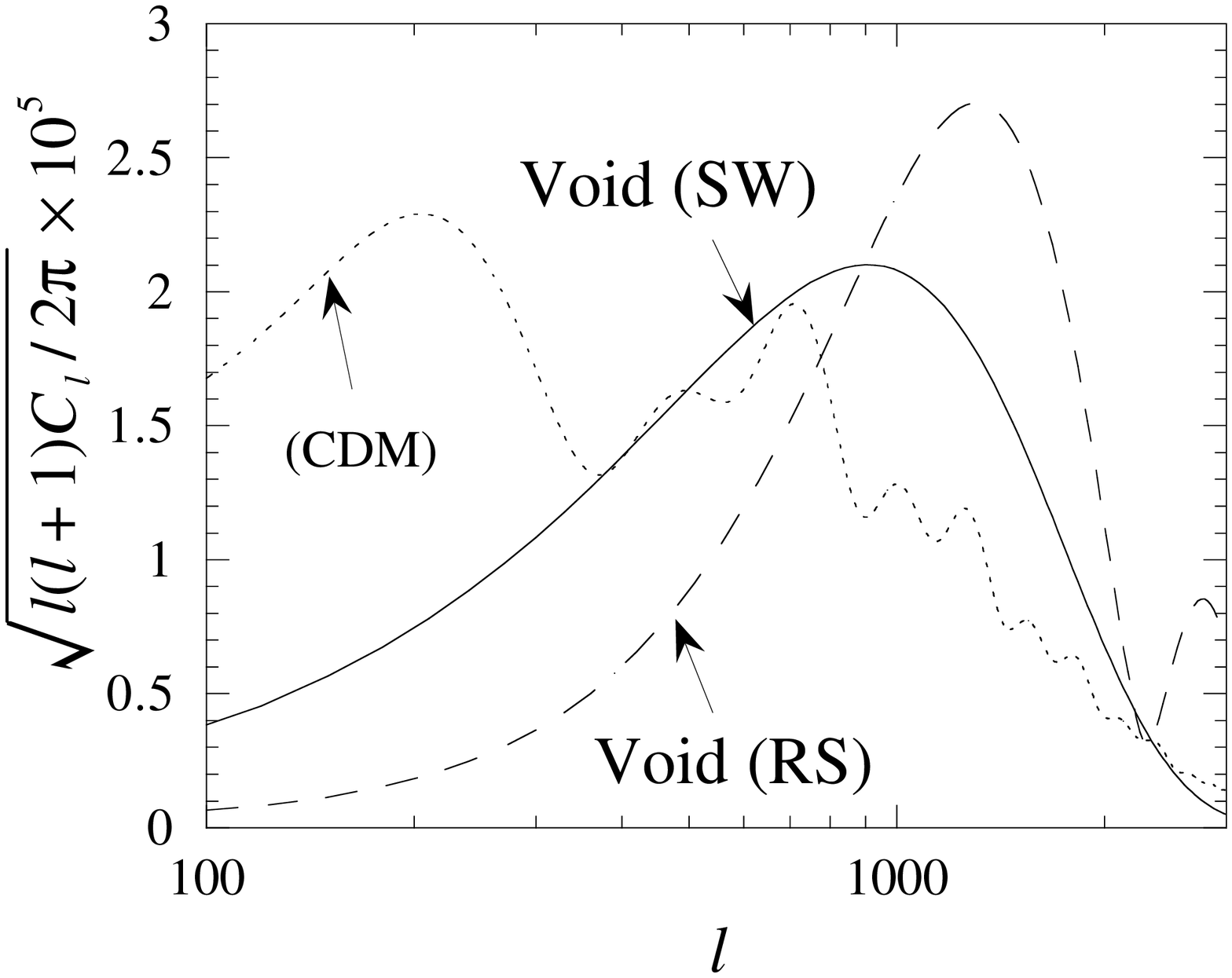,width=12cm}
\figcaption{RS effect and SW effect: plots of the CMB angular power spectrum 
$\sqrt{l(l+1)C_l}$ v.s. $l$. (a), (b), and (c) correspond to Models (a), (b), and (c), respectively. $z_f=1000$. 
\label{fig8}}

\epsfile{file=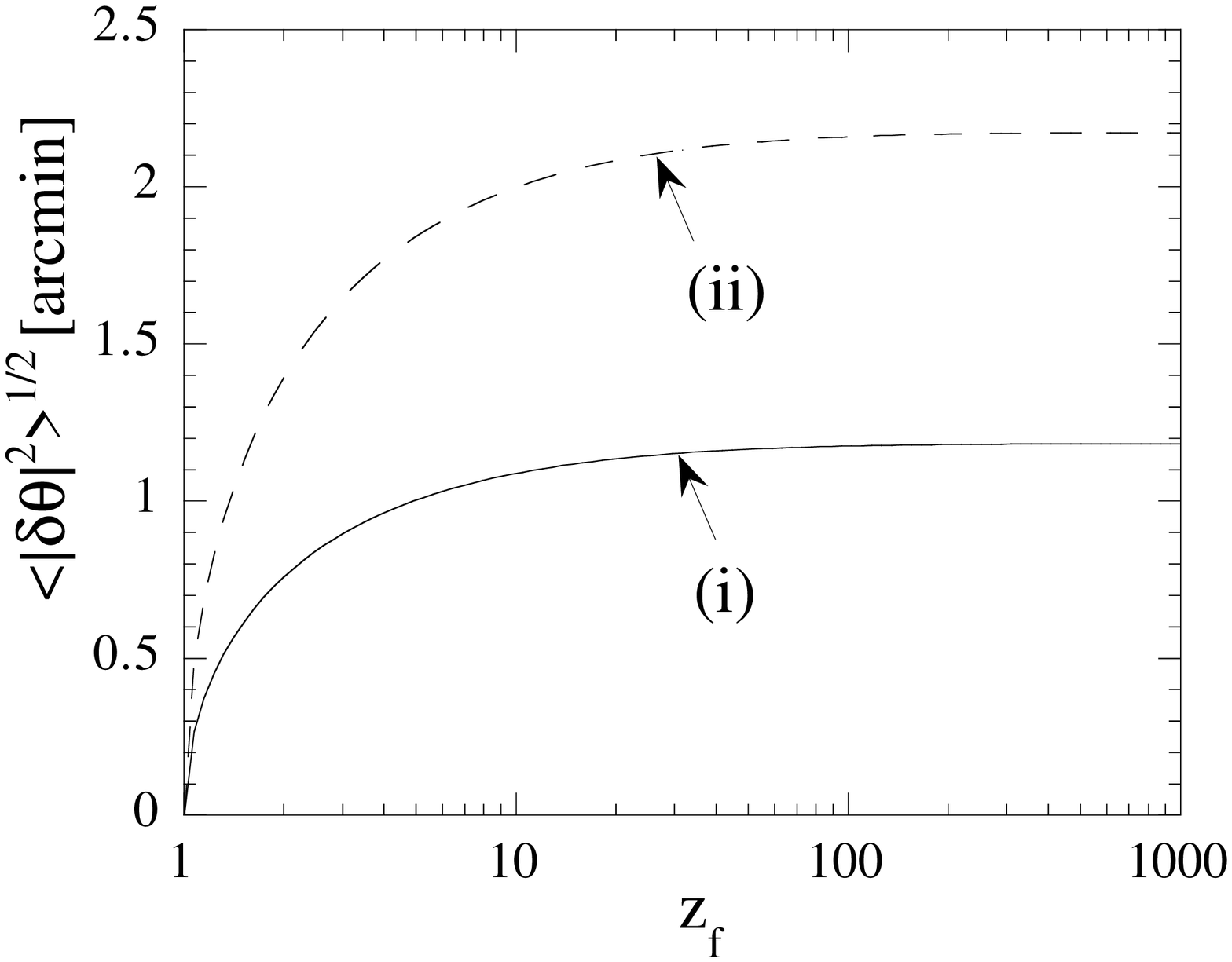,width=12cm}
\figcaption{Lensing effect: plot of $\sqrt{<|\delta\bt|^2>}$ v.s. $z_{{\rm f}}$. \label{fig9}}

\vskip 5mm
\epsfile{file=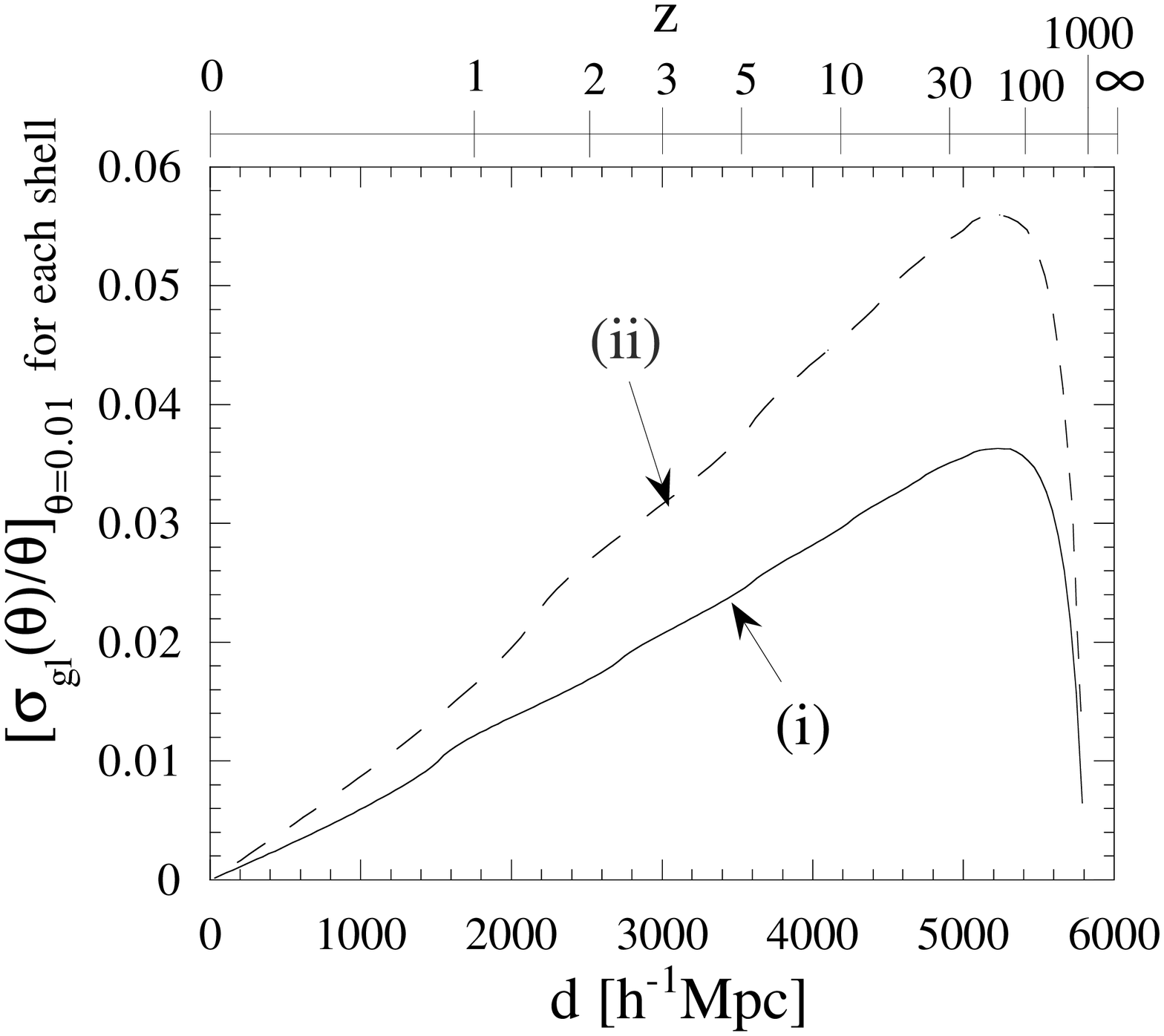,width=12cm}
\figcaption{Lensing effect: contribution by voids in each shell. The
abscissa is a position of shells in terms of $h^{-1}$Mpc and $z$.
\label{fig10}}

\epsfile{file=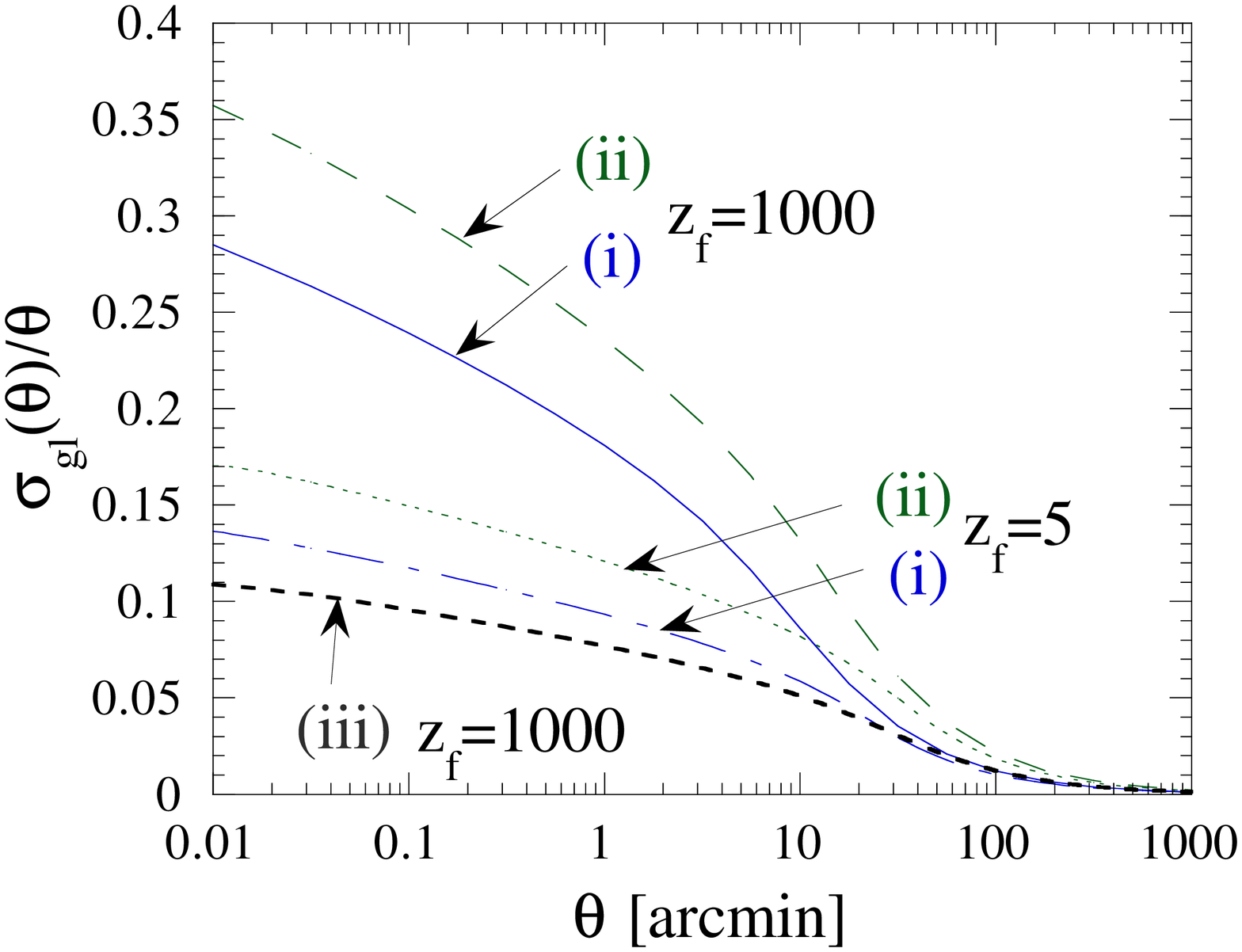,width=12cm}
\figcaption{Lensing effect: plots of $\sigma_{{\rm gl}}(\theta)/\theta$ v.s. $\theta$. We show five examples: Model (i)
with $z_f=5$, (i) with $z_f=1000$, (ii) with $z_f=5$, (ii) with $z_f=1000$, and (iii) with $z_f=1000$, 
\label{fig11}}

\vskip 5mm
(a) \epsfile{file=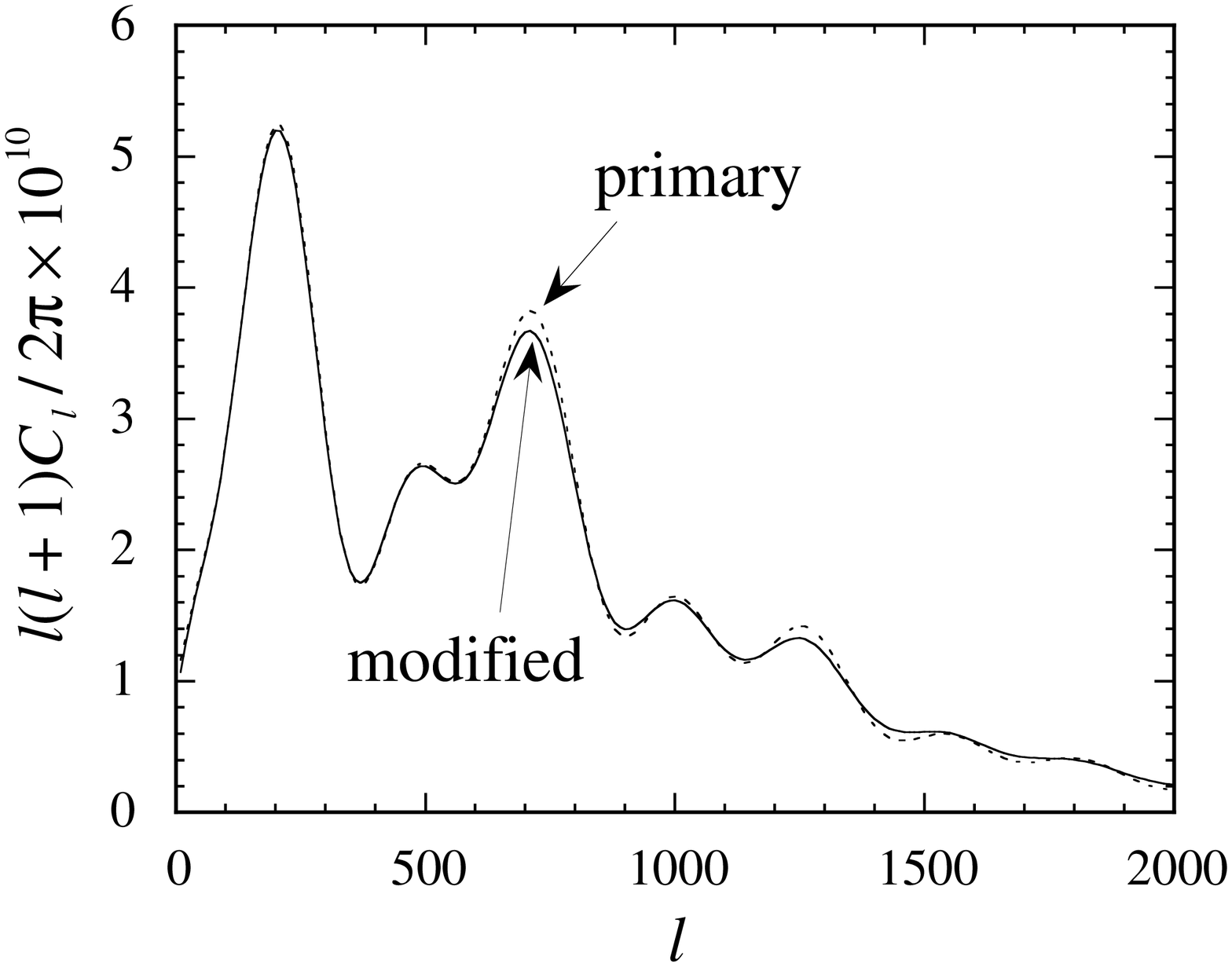,width=12cm}

(b) \epsfile{file=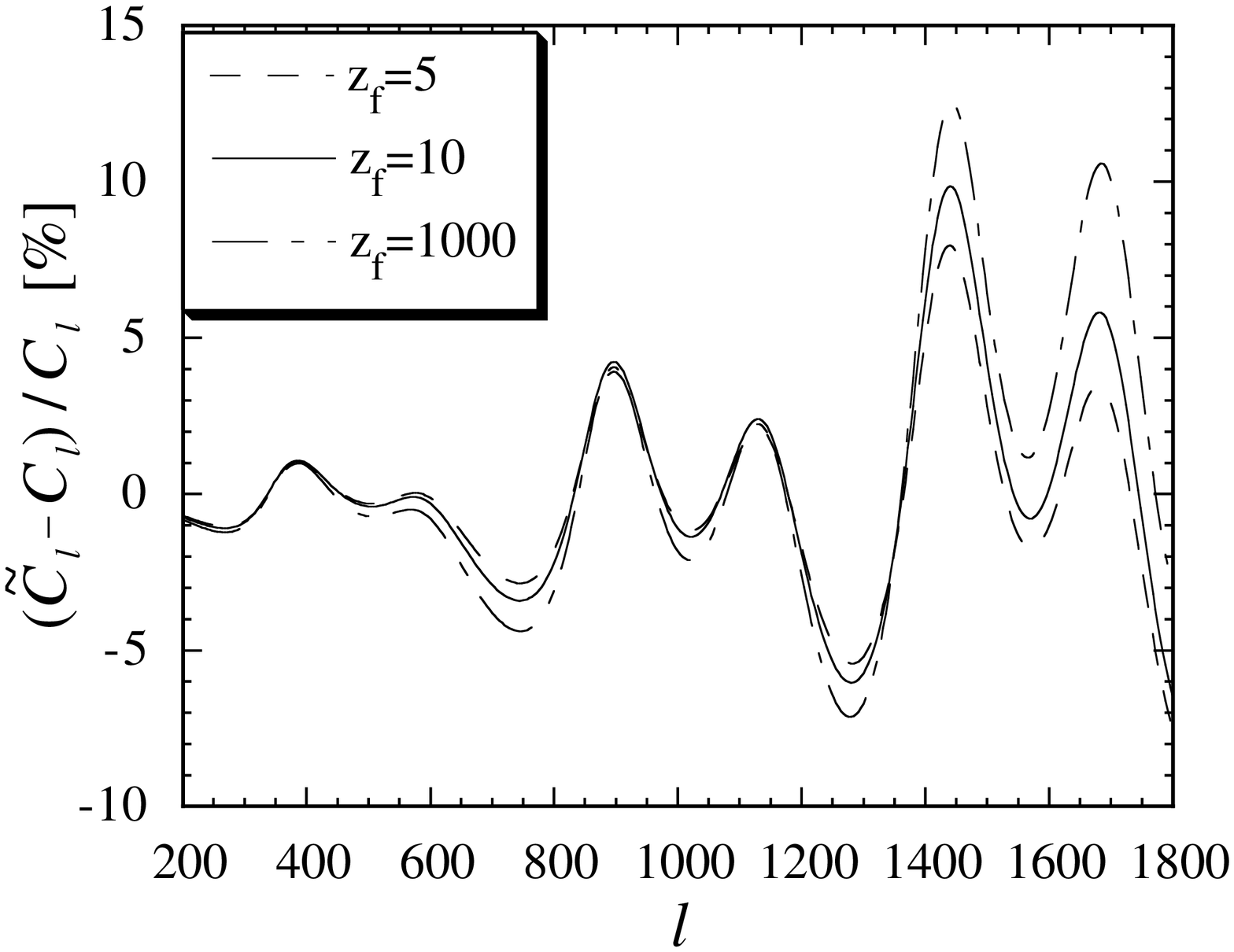,width=12cm}

\vskip 5mm
(c) \epsfile{file=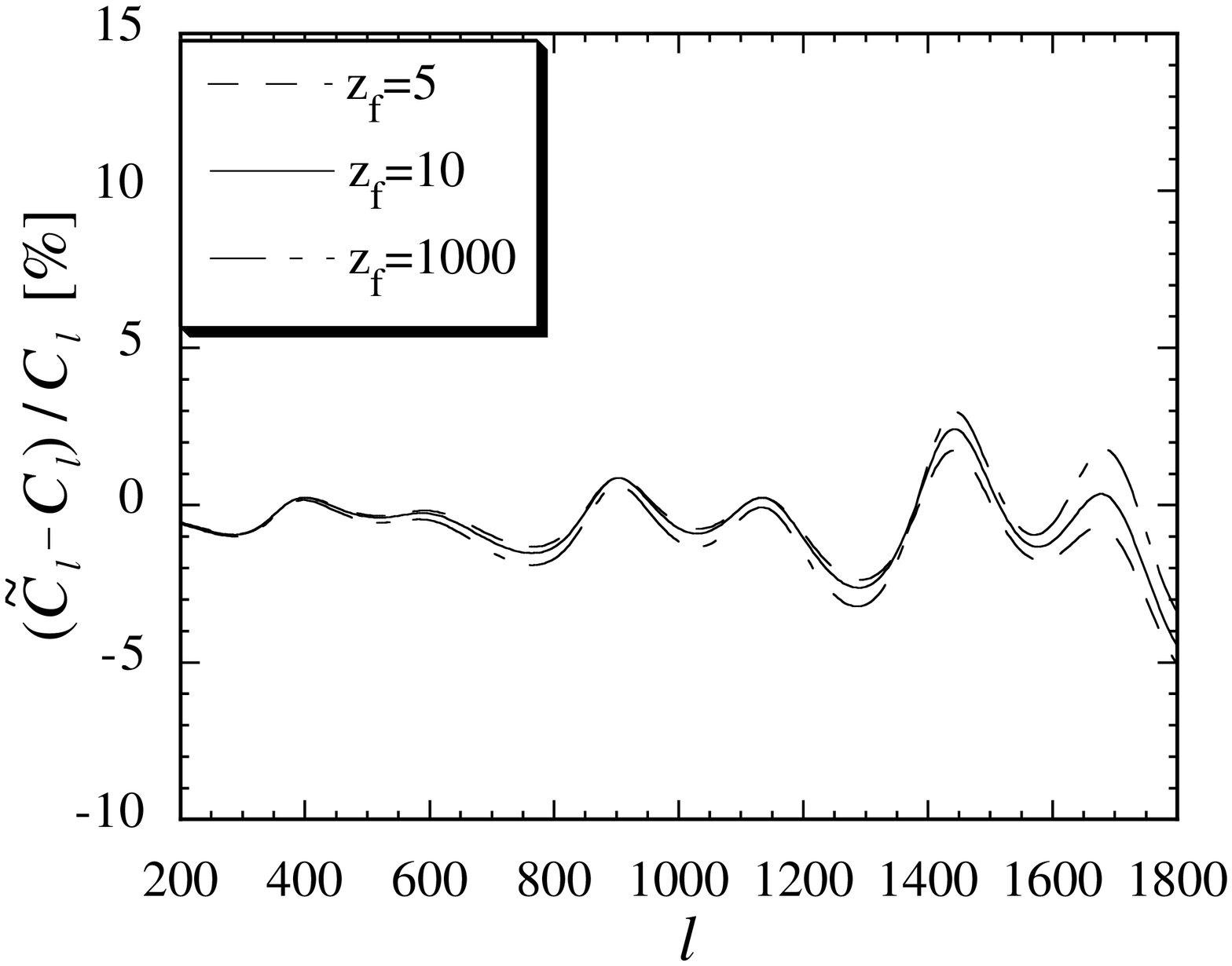,width=12cm}
\figcaption{Lensing effect. (a) shows plots of the CMB angular power spectrum $l(l+1)C_l$ v.s. $l$ with lensing
and without lensing for Model (ii). In (b) and in (c), respectively, the relative
changes of the spectrum due to lensing for Model (i) and for Model (ii)
are presented. \label{fig12}}

\end{document}